\documentclass{iau}

\usepackage{amsmath}
\usepackage{graphicx}
\usepackage{multirow}

\begin{document}

\lefttitle{Eggl et al.}
\righttitle{Satellite Interference Around the World}

\jnlPage{x}{x}
\jnlDoiYr{2023}
\doival{10.1017/xxxxx}

\aopheadtitle{Proceedings IAU Symposium}
\journaltitle{Astronomy and Satellite Constellations: Pathways Forward}
\editors{P. Grimley, eds.}
\volno{385}  %% insert here IAU Symposium No.

\title{SatHub Panel: Satellite Interference in Observatories Around the World}

\author{Siegfried~Eggl$^{1,11}$, Zouhair~Benkhaldoun$^2$,  Genoveva~Micheva$^3$, Samuel~T.~Spencer$^{4,5}$, David~V.~Stark$^{6,7}$, Benjamin~Winkel$^8$, Meredith~Rawls$^{9,11}$, Mike~W.~Peel$^{10,11}$}
% \affiliation{Department of Aerospace Engineering, University of Illinois at Urbana-Champaign,
% 61801 IL, Champaign, USA, email: {\tt eggl@illinois.edu}}
\affiliation{$^1$ Department of Aerospace Engineering, University of Illinois at Urbana-Champaign,
61801 IL, Champaign, USA \\ email: {\tt eggl@illinois.edu}}
\affiliation{$^2$ Oukaimeden Observatory, High Energy Physics and Astrophysics Laboratory, Cadi Ayyad University, \\ Marrakech, Morocco \\ email: {\tt zouhair@uca.ac.ma}}
\affiliation{$^3$ Leibniz-Institute for Astrophysics Potsdam (AIP) \\ An der Sternwarte 16, 14482 Potsdam, Germany \\email: {\tt gmicheva@aip.de}}
\affiliation{$^4$ Friedrich-Alexander-Universit{\"a}t Erlangen-N{\"u}rnberg\\ Erlangen Centre for Astroparticle Physics\\ Nikolaus-Fiebiger-Str. 2, D 91058 Erlangen, Germany \\email: {\tt samuel.spencer@fau.de}}
\affiliation{$^5$ Department of Physics \\ Clarendon Laboratory, Parks Road, Oxford, OX1 3PU, United Kingdom}
\affiliation{$^6$ Space Telescope Science Institute \\ 3700 San Martin Dr, Baltimore, MD 21218, USA \\email: {\tt dstark@stsci.edu}}
\affiliation{$^7$ Department of Physics \& Astronomy, Johns Hopkins University, \\ 3400 N. Charles Street, Baltimore, MD 21218, USA}
\affiliation{$^8$ Max-Planck Institute for Radioastronomy, \\ Bonn, Germany  \\email: {\tt bwinkel@mpifr-bonn.mpg.de}}
\affiliation{$^9$ Department of Astronomy/DiRAC/Vera C. Rubin Observatory, University of Washington, Seattle, Washington, USA. \\ email: {\tt mrawls@uw.edu}}
\affiliation{$^{10}$ Imperial College London, Blackett Lab, Prince Consort Road, London SW7 2AZ, UK  \\ email: {\tt email@mikepeel.net}}
\affiliation{$^{11}$ IAU Centre for the Protection of the Dark and Quiet Sky from Satellite Constellation Interference, \\ Paris, France \\email: {\tt sathub@cps.iau.org}}

\begin{abstract}
Satellite constellation interference occurs across astronomical disciplines. We present examples of interference from radio and $\gamma$-Ray astronomy to optical and spectroscopic interference in ground-based and space-borne facilities. In particular, we discuss the impact of artificial satellites on  the Hubble Space Telescope (HST), the High Energy Stereoscopic System (H.E.S.S.), an Imaging Atmospheric Cherenkov Telescope, as well as possible mitigation strategies for the European Southern Observatory 4-metre Multi-Object Spectrograph Telescope (ESO 4MOST). Furthermore, we shed light on how ground-based optical telescopes such as the Oukaimeden Observatory contribute to IAU Centre for the Protection of the Dark and Quiet Sky from Satellite Constellation Interference (IAU CPS) efforts that quantify satellite brightness.  
%% add here a maximum of 10 keywords, to be taken form the file <Keywords.txt>
\end{abstract}


\maketitle

% \firstsection % if your document starts with a section,
%               % remove some space above using this command.
\section{Introduction}
Satellite constellation interference with ground-based optical and radio astronomy facilities is well documented \citep[e.g.,][]{barentine_aggregate_2023, karpov_rate_2023, nandakumar_high_2023,anderson-baldwin_leo_2023, grigg_detection_2023}. The issue of satellites passing through the field of view (FoV) of observatories and potentially corrupting astronomical data is not limited to those domains, however. In this contribution we discuss examples of satellite interference across astronomical disciplines. After a brief overview of challenges and potential mitigation strategies in Radio Astronomy (RA) in section \ref{sec:RA} and ground-based 
 spectroscopy in section \ref{sec:4MOST} we discuss the impact of satellite constellations on $\gamma$-ray astronomy and space-based facilities in sections \ref{sec:XRAY} and \ref{sec:HST}. We conclude this article with an example of how ground-based observatories can contribute to IAU Centre for the Protection of the Dark and Quiet Sky from Satellite Constellation Interference \citep{CPS2023} efforts to quantify satellite interference with ground based astronomy in section \ref{sec:OK}.

% Winkel
\section{A Radio Astronomer's Perspective}
\label{sec:RA}
Large satellite constellations impact modern radio astronomy (RA) in various ways. Classically, impact assessment studies have focused on intentional transmissions and their by-products (e.g. spectral side-lobes) in the bands allocated and adjacent to the satellite radio services \citep{di2023large}. Recently, however, other interference vectors have emerged, such as electromagnetic leakage from onboard electronics \citep{di_vruno_unintended_2023,grigg_detection_2023}. Some satellite constellation and terrestrial cell-phone network operators are also teaming up to smoothly integrate both systems. This provides extra challenges to RA as the carefully constructed coordination scheme between cell phone networks and the affected radio services does not include space-based infrastructure. 
In order to enable a productive co-existence of RA and satellite constellations coordination and regulatory action are required. All regulatory activities at the International Telecommunication Union (ITU-R) and other organisations are based on compatibility studies.
Spectrum managers are often asked by other stakeholders whether they can simply ``try things out", i.e. assess interference based on real-life testing of assets. 
That may not always be possible, since RA protection criteria are extremely strict: after about an hour of integration, any interference must still be below 10\% of the noise level. That means observational proof that protection levels are violated can require an extraordinary amount of observation time. Moreover, some observatories may not be capable of tracking artificial satellites.
Modeling is sometimes the only viable route to determining potential satellite interference with RA facilities.
As large constellations can consist of thousands of satellites, full network simulations are required. Such simulations have to consider the dynamic behavior of transmitters and receivers, antenna patterns (including side-lobes and time-variable pointing), aggregation effects and - at higher frequencies - atmospheric effects. The International Telecommunication Union has developed procedures and algorithms that can assist such calculations. Nevertheless, the implementation is non-trivial, especially with respect to computational costs.
The use of official algorithms and models provided by the ITU-R is strongly recommended in order to avoid unnecessary ambiguities.
Observations are still required to ensure calibration of the model computations as well as to determine technical parameters of satellites, which operators may be hesitant to share publicly.
An important point here is that observations should produce results that are calibrated with respect to ITU-R standard models. Detection experiments alone may not be compatible with and/or useful for regulatory purposes.

% Micheva
\section{Satellite Constellation Interference and Mitigation in Spectroscopy}
\label{sec:4MOST} 
Active avoidance, i.e. pointing the telescope ``somewhere else" to avoid FoV crossings of satellites, is often proposed as a satellite interference mitigation strategy. Unfortunately, this strategy is not always feasible or economical \citep[e.g.,][]{hu2022satellite}.
This is also the case for the European Southern Observatory 4-metre Multi-Object Spectroscopic Telescope (4MOST), which is a state-of-the-art spectroscopic survey instrument being developed for the ESO VISTA telescope. It has a wide 4.2 square degree FoV, populated by 2436 optical fibers which can capture celestial objects simultaneously. While the wide FoV makes 4MOST suitable for extensive and efficient surveys of the southern sky, the extended exposure durations make 4MOST vulnerable to contamination by satellites crossing the wide FoV. Strategies need to be developed to identify potential corruption in 4MOST spectroscopic data. On average, each satellite that traverses the 4MOST FoV will affect approximately 1.3 fibers. While many of these satellite trails will be too dim to impact the 4MOST spectra significantly, some will be bright enough to subtly alter the data without clear indication that something went wrong. Without additional mitigation, this issue would make it impossible to rely on individual 4MOST spectra; only large statistical samples could be safely analyzed. This would, for instance, negatively affect the exploration of anomalous spectra for groundbreaking new scientific discoveries. Simulations with 4MOST indicate that satellite contamination can also subtly affect stellar metrics like radial velocity and elemental abundances, depending on the satellite's brightness level. These effects are so subtle that they evade detection by existing data processing and quality control pipelines. To address this, 4MOST plans to employ a specialized imager for real-time, continuous FoV monitoring to identify crossing satellites. However, the technical specifications for this imager present challenges that have yet to be resolved and add substantial complexity to the project.

% Samuel T. Spencer
\section{Satellite Constellation Interference in $\gamma$-Ray Astronomy}
\label{sec:XRAY}
 High Energy Stereoscopic System (H.E.S.S.) is an Imaging Atmospheric Cherenkov Telescope (IACT) array located in Namibia, and is designed to perform ground-based $\gamma$-ray astronomical observations. It works by detecting the optical to near-Ultra Violet Cherenkov light emitted when $\gamma$-ray photons interact with Earth’s atmosphere and initiate a particle cascade, a so-called ``Extensive Air Shower" (EAS). Unlike in optical or radio astronomy, the presence of very-high-energy $\gamma$-rays is inferred by classifying and determining the properties of these EAS events. The effect of satellite trails on these IACT observations has previously been considered minimal, due to their approximately 10 nanosecond signal integration time. However, IACTs are still exposed to the night sky and sensitive to optical frequency light, and therefore can be affected by satellites. Given the continuing satellite launches, and the construction of the next-generation Cherenkov Telescope Array (CTA) observatory, it is necessary to quantify such interference. CTA construction efforts have already begun on La Palma with the inauguration of Large Sized Telescope 1 (LST-1); a second site for CTA will be located in Chile (near Cerro Paranal). \cite{lang2023impact} developed a method of detecting satellites in H.E.S.S. data using night sky monitoring measurements, which H.E.S.S. cameras take concurrently with $\gamma$-ray observations. \cite{lang2023impact} show that, whilst measurable, the effect of satellite trails on EAS event classification and reconstruction with H.E.S.S. is minimal. However, a mildly increasing trend in the number of trails was detected over the past three years, and new analysis techniques that aim to reduce the energy threshold of IACT observations with next generation instruments could be more severely affected.
 
% David V. Stark
\section{Satellite Constellation Interference with Space-borne Facilities}
\label{sec:HST}
Space-borne telescopes in low-Earth orbit (LEO) can also be affected by satellite streaks. To demonstrate this, \cite{stark2022improved} studied 30,000 images taken between 2002 and mid-2022 using the Wide-Field-Channel (WFC) on the Advanced Camera for Surveys (ACS) aboard the Hubble Space Telescope (HST). \cite{stark2022improved} find that the rate of satellite trails in ACS/WFC imaging data has increased by approximately a factor of two in the last two decades, with approximately 10\% of images currently affected, but there is no clear systematic evolution in the typical trail brightness.
The Space Telescope Science Institute has spear-headed the development of a new software package that was used to identify and mask satellite trails in HST astronomical imaging data \citep{stark2022improved}. The approach is based around the Median Radon Transform, a modified version of the standard Radon Transform. The key advantage of this method is its sensitivity; it can detect linear features with mean brightness significantly below the background noise level of an image, and it is resistant to false detections caused by bright astronomical sources (e.g., stars, galaxies) in the majority of cases. The software has been incorporated into the “acstools” Python package that is publicly available.

% Benkhaldoun
\section{What can astronomers do?}
\label{sec:OK}
 The administration of space traffic is getting increasingly complex, needing growing participation from both professional and amateur astronomers. The Cadi Ayad University Observatory, in partnership with the IAU CPS and SpaceAble, is actively involved in satellite and space debris observation. 
 The IAU CPS is an international organization dedicated to working with all stakeholders to tackle the challenges artificial satellite constellations pose to astronomy. Participation in IAU CPS observing campaigns allows observatories around the world to contribute, analyze and publish satellite brightness data, for instance, on AST SpaceMobile's BlueWalker 3 \citep{nandakumar_high_2023} and several generations of Starlink satellites \citep{krantz_steward_2023,mallama_assessment_2023}. The Oukaimeden observatory is a partner of the IAU CPS center and as such, the observatory's researchers are directly involved in the coordination and acquisition of satellite observations through the deployment of a variety of professional and amateur telescopes. The observatory also provides facilities for accommodating and collaborating with other satellite observation programs, as well as an academic and professional training program to support observational activities. Furthermore, utilizing orbital data processing techniques, the observatory has developed a variety of unique platforms that aid in observation planning. These user-friendly interfaces enable more effective observation scheduling, thereby increasing the observatory's contribution in the various scientific objectives behind satellite observations.
 The extensive activities carried out by the Oukaimeden observatory in coordination with the IAU CPS include observation planning, satellite tracking, and the extraction of orbital parameters through image processing. These efforts are an excellent example of how astronomers can contribute to assessing the impact of satellite constellations on ground based-astronomy.
\section{Acknowledgements}
STS is supported by the Deutsche Forschungsgemeinschaft (DFG, German Research Foundation) – Project Number 452934793. SE and MP thank the IAU CPS for assisting with travel support. 
\bibliography{references}
\bibliographystyle{iaulike}

\end{document}